# Efficient spin transport in a paramagnetic insulator

Koichi Oyanagi[1], Saburo Takahashi[1,2,3], Ludo J. Cornelissen[4], Juan Shan[4], Shunsuke Daimon[1,2,5], Takashi Kikkawa[1,2], Gerrit E. W. Bauer[1,2,3,4], Bart J. van Wees[4], and Eiji Saitoh[1,2,3,5,6]

1. Institute for Materials Research, Tohoku University, Sendai 980-8577, Japan

2. Advanced Institute for Materials Research, Tohoku University, Sendai 980-8577, Japan

3. Center for Spintronics Research Network, Tohoku University, Sendai 980-8577, Japan

4. Physics of Nanodevices, Zernike Institute for Advanced Materials, University of Groningen, Nijenborgh 4, 9747 AG Groningen, The Netherlands

5. Department of Applied Physics, The University of Tokyo, Tokyo 113-8656, Japan

6. Advanced Science Research Center, Japan Atomic Energy Agency, Tokai 319-1195, Japan

**The discovery of new materials that efficiently transmit spin currents has been important for spintronics and material science[1-5]. The electric insulator $Gd_3Ga_5O_{12}$ (GGG) is a superior substrate for growing magnetic films[6], but has never been considered as a conduit for spin currents. Here we report spin current propagation in paramagnetic GGG over several microns. Surprisingly, the spin transport persists up to temperatures of 100 K >> $T_g$ = 180 mK, GGG's magnetic glass-like transition temperature[7]. At 5 K we find a spin diffusion length $\lambda_{GGG}$ = 1.8 ± 0.2 μm and a spin conductivity $\sigma_{GGG}$ = (7.3 ± 0.3) × $10^4$ $Sm^{-1}$ that is larger than that of the record quality magnet $Y_3Fe_5O_{12}$ (YIG). We conclude that exchange coupling is not required for efficient spin transport, which challenges conventional models and provides new material-design strategies for spintronic devices.**

According to conventional wisdom, spin currents can be carried by conduction electrons and spin waves. Mobile electrons in a metal carry spin currents over distances typically less than a micron, while spin waves, the collective excitation of the magnetic order parameter[8-15] can communicate spin information over much longer distances (Fig. 1a). In particular, the ferrimagnetic insulator $Y_3Fe_5O_{12}$ (YIG) supports spin transport over up to a millimeter[1,2].

$Gd_3Ga_5O_{12}$ (GGG) is an excellent substrate material for the growth of, e.g., high quality YIG films[6]. Above $T_g$ = 180 mK, it is a paramagnetic insulator (band gap of 6 eV[16]) with $Gd^{3+}$ spin-7/2 local magnetic moments that are weakly coupled by an effective spin interaction[17] $J_{ex}$ ~ 100 mK. Hence, GGG does not exhibit long-range magnetic ordering at all temperatures[7] (Fig. 1d), while its field-dependent magnetization is well described by the Brillouin function. The low-temperature saturation magnetization of ~ 7 $\mu_B/Gd^{3+}$ (Fig. 1c) is governed by the half-filled 4$f$-shell of the $Gd^{3+}$ local moments and larger than that of ferrimagnetic YIG (~ 5 $\mu_B/Fe^{3+}$).

Paramagnetic insulators have not attracted the spintronics community's attention since they seem unlikely carriers of spin currents. The charge gap prohibits electron conduction, and in the absence of a substantial exchange stiffness coherent spin waves are not expected to exist. Few papers address the spin transport properties of paramagnets[18-23]. Here we report nonlocal spin transport over microns in a GGG slab (Fig. 1b) even at elevated temperature, implying a surprisingly good spin conductivity. At low temperatures, GGG turns out to be a better spin-conductor even than YIG.

We adopt the standard nonlocal geometry[8-15] to study spin transport in a device comprised by two Pt wires separated by a distance $d$ on top of a GGG slab (Fig. 2b). Here, spin currents are injected and detected via the direct and inverse spin Hall effects[1,25] (SHE and ISHE), respectively (Fig. 2a). A charge current, $J_c$, in one Pt wire (injector) generates non-equilibrium spin accumulation $\mu_s$ with direction $\sigma_s$ at the Pt/GGG interface by the SHE. When $\sigma_s$ and the magnetization **M** in GGG are collinear, the interface spin-exchange interaction transfers spin angular momentum from the conduction electron spins in Pt to the local moments in GGG at the interface, thereby creating a non-equilibrium magnetization in the GGG beneath the contact that generates a spin diffusion current into the paramagnet. Some of it will reach the other Pt contact (detector) and generate a transverse voltage in Pt by means of spin pumping into Pt and the ISHE.

To our knowledge, long-range spin transport in insulators has been observed only in magnets below their Curie temperaturs[8-15], e.g. YIG, NiFe$_2$O$_4$, and α-Fe$_2$O$_3$, so magnetic order and spin-wave stiffness have been considered indispensable. Here, we demonstrate that a relatively weak magnetic field can be a sufficient condition for efficient spin transport, and assistance by coherent spin waves is not required.

First, we discuss the field and temperature dependence of the nonlocal detector voltage $V$ in Pt/GGG/Pt with contacts at a distance $d = 0.5$ μm. We use the standard lock-in technique to rule out thermal effects (see Methods). A magnetic field $B$ is applied at angle $\theta = 0$ in the $z$-$y$ plane (see Fig. 2b) such that the magnetization in GGG is parallel to the spin polarization $\boldsymbol{\sigma}_s$ of the SHE-induced spin accumulation in the injector.

Figure 2c shows $V(B)$ at 300 and 5 K. Surprising is the voltage observed at low, but not ultralow temperatures that increases monotonically as a function of $|B|$ and saturates at about 4 T. Pt/YAG/Pt, where YAG is the diamagnetic insulator $Y_3Al_5O_{12}$, does not generate such a signal (Fig. 3b); apparently the paramagnetism of GGG (Fig. 1c) is instrumental for the effect.

We present the nonlocal voltage in Pt/GGG/Pt at 5 K as a function of the out-of-plane magnetic-field angle $\theta$ and injection-current $J_c$ as defined in Fig. 2b. The left panel of Fig. 2d shows that $V$ at $|B| = 3.5$ T. $V(\theta) = V_{\max} \cos^2 \theta$: it is maximal ($V_{\max}$) at $\theta = 0$ and $\theta = \pm$ 180° (**B** // **y**) but vanishes at $\theta = \pm 90°$ (**B** // **z**). Furthermore, $V$ depends linearly on $|J_c|$ (see the right panel of Fig. 2d). The same $\cos^2\theta$ dependence in Pt/YIG/Pt is known to be caused by the injection and detection efficiencies of the magnon spin current by the SHE and ISHE, respectively, that both scale like $\cos\theta$ (refs 8, 9). The SHE-induced spin accumulation at the Pt injector interface, i.e. the driving force for the nonlocal magnon transport, scales linearly with $J_c$, and so does the nonlocal $V$ (ref. 8). The above observations are strong evidence for spin-current transport in paramagnetic GGG without long-range magnetic order.

Figure 3c shows the $\theta$ dependence of $V$ at $B = 3.5$ T and various temperatures $T$. Clearly $V \sim V_{max}(T)\cos^2\theta$, where $V_{max}(T)$ in Fig. 3a decreases monotonically for $T > 5$ K, nearly proportional to $M(T)$ at the same $B$ as shown in the inset to Fig. 3a. The field-induced (para)magnetism therefore plays an important role in the $|V|$ generation. Surprisingly, $V_{max}$ at 3.5 T persists even at 100 K, which is two orders of magnitude larger than the Curie-Weiss temperature $|\Theta_{CW}| = 2$ K. The exchange interaction at those temperatures can therefore not play any role in the voltage generation. In contrast, a large paramagnetic magnetization ($\sim \mu_B$ at 3.5 T) is still observed at 100 K, consistent with long-range spin transport carried by the field-induced paramagnetism.

At high fields, Fig. 3d shows a non-monotonic $V(B, \theta = 0)$: At $T < 30$ K, $V$ gradually decreases with field after a maximum at ~ 4 T, which becomes more prominent with decreasing $T$. A similar feature has been reported in Pt/YIG/Pt[12] and interpreted in terms of the freeze-out of magnons: a Zeeman gap $\propto B$ larger than the thermal energy $\propto T$ critically reduces the magnon number and conductivity. It appears that thermal activation of magnetic fluctuations is required to enable a spin current in GGG as well.

By changing the distance $d$ between the Pt contacts, we can measure the penetration depth of an injected spin current. $V_{max}$ at 5 K, as plotted in Fig. 4a, decreases monotonously with increasing $d$. A similar dependence in Pt/YIG/Pt is well described by a magnon diffusion model[25,26]. We postulate that the observed spin transport in GGG can be described in terms of incoherent (over-damped) spin transport that obeys the same phenomenology as magnon diffusion. Since the GGG thickness of 500 μm >> $d$, we cannot use a simple one-dimensional diffusion model, which predicts a simple exponential decay $V_{max}(d) \sim \exp(-d/\lambda)$. We rather have to consort to a diffusion model in two spatial dimensions (see Section D of

Supplementary Information) which leads to:

$$V_{\text{max}}(d) = CK_0(d/\lambda), \quad (1)$$

where $K_0(d/\lambda)$ is the modified Bessel function of the second kind, $\lambda = \sqrt{D\tau}$ is the spin diffusion (relaxation) length, $D$ is the diffusion constant, $\tau$ is the relaxation time, and $C$ is a numerical coefficient that does not depend on $d$. By fitting equation (1) to the experimental data, we obtained $\lambda_{\text{GGG}} = 1.82 \pm 0.19$ μm at $B = 3.5$ T and $T = 5$ K.

A long spin relaxation length in an insulator implies weak spin-lattice relaxation by spin-orbit interaction, which in GGG should be weak because the 4$f$-shell in Gd$^{3+}$ is half-filled with zero orbital angular momentum ($L = 0$). We tested this scenario by a control experiment on a Pt/Tb$_3$Ga$_5$O$_{12}$ (TGG)/Pt nonlocal device with similar geometry ($d = 0.5$ μm). TGG is also a paramagnetic insulator with large field-induced $M$ at low temperatures (see the inset to Fig. 3a). However, Tb$^{3+}$ ions have a finite orbital angular momentum ($L = 3$) and electric quadrupole that strongly couple to the lattice[27]. Indeed, we could not observe any nonlocal voltage in Pt/TGG/Pt in the entire temperature range (see Fig. 3a,b). This result highlights the importance of a weak spin-lattice coupling in the long-range paramagnetic spin transmission.

We model the nonlocal voltages in a normal-metal (N)/paramagnetic-insulator (PI)/normal-metal (N) system by the spin diffusion equation in the PI with proper boundary conditions at the cantacts[28], the spin diffusion, and spin-charge conversion in the metal (see Sections C to F of Supplementary Information). The voltage in the Pt detector as a function of $B$ and $T$ reads:

$$V(B, T) = C_1 \frac{g_s^2}{\sigma_{GGG}} \frac{[\xi B_S(2S\xi)/\sinh(\xi)]^2}{[1 + 2\xi C_2 \frac{g_s}{\sigma_{GGG}} B_S(2S\xi)]^2}, \qquad (2)$$

where $\xi(B,T) = g\mu_B B / k_B(T + |\Theta_{CW}|)$, $g$ is the $g$-factor, $\mu_B$ is the Bohr magneton, $k_B$ is the Boltzmann constant, $B_S(x)$ is the Brillouin function as a function of $x$ for spin-$S$, $C_1$ and $C_2$ are known numerical constants, $S = 7/2$ is the electron spin of a $Gd^{3+}$ ion, $g_s$ is the effective spin conductance of the Pt/GGG interface, and $\sigma_{GGG}$ is the spin conductivity in GGG. The observed $V(B)$ is well described by equation (2) (see Fig. 4b). The best fit of equation (2) is achieved by $\sigma_{GGG} = (7.25 \pm 0.26) \times 10^4$ Sm$^{-1}$ and $g_s = (1.82 \pm 0.05) \times 10^{11}$ Sm$^{-2}$. We also determined $\sigma_{GGG}$ and $g_s$ by a numerical (finite-element) simulation of the diffusion[26] that takes the finite width of the contacts and the GGG film into account and find good agreement with the analytical model (the details are discussed in Supplementary Information Section G).

Surprisingly, both the obtained $g_s$ and $\sigma_{GGG}$ values are, at the same temperature, six times larger than those of the Pt/YIG/Pt sample[11], which is evidence for highly efficient paramagnetic spin transport.

Finally, we discuss the mechanism of the spin transport in GGG. At low temperatures, only low-frequency spin waves close to $k = 0$ can contribute to the spin current at which the exchange interaction between the local moments can be disregarded since it scales like $k^2$. The spin-wave dispersion and transport are therefore dominated by magnetic dipole-dipole interaction. Rather than a help, exchange interaction can be a nuisance, since short-ranged and sensitive to local magnetic disorder such as grain boundaries that give rise to spin-wave scattering. In contrast, dipole interaction is long-ranged and less sensitive to disorder. Therefore, paramagnets which exhibit strong dipole coupling free from exchange interaction, such as GGG, may provide ideal spin conductors when thermal fluctuation of the magnetic moments is sufficiently suppressed by applied magnetic fields. Finally, the large spin conductance $g_s$ indicates that the interface exchange interaction between a metal and GGG is not suppressed as compared to YIG.

In summary, we discovered long-range spin transport in the Curie-like paramagnetic insulator $Gd_3Ga_5O_{12}$. Spin current propagates over several microns and its transport efficiency at moderately low temperatures is even higher than that of the best magnetically ordered material YIG. Since paramagnetic insulators are free from magnetic-domain Barkhausen noise and aging (magnetic after-effects)[29] typical for ferromagnets and antiferromagnets, they are promising materials for future spintronics devices.

**Methods**

**Sample preparation**

A single-crystalline $Gd_3Ga_5O_{12}$ (111), $Y_3Al_5O_{12}$ (111), and $Tb_3Ga_5O_{12}$ (111) (500 μm in thickness) were commercially obtained from CRYSTAL GmbH, Surface Pro GmbH, and MTI corporation, respectively. For magnetization measurements, the slabs were cut into 3 mm long and 2 mm wide. For transport measurements, a nonlocal device with Pt wires was fabricated on a top of each slab by an e-beam lithography and lift-off technique. Here, the Pt wires were deposited by magnetron sputtering in a $10^{-1}$ Pa Ar atmosphere. The dimension of the Pt wire is 200 μm long, 100 nm wide, and 10 nm thick and the separation distance between the injector and detector Pt wires are ranged from 0.3 to 3.0 μm. A microscope image of a device is presented in Supplementary Information Section A. We measured the temperature dependence of the resistance between the two Pt wires on the GGG substrate but found that the resistance of the GGG is too high to be measurable.

**Magnetization measurement**

The magnetization of GGG, YAG, and TGG slabs was measured using a Vibrating Sample Magnetometer (VSM) option of a Quantum Design Physical Properties Measurement System (PPMS) in a temperature range from 5 K to 300 K under external magnetic fields up to 9 T.

**Spin transport measurement**

Spin transport measurements were carried out by a standard lock-in technique with a PPMS in a temperature range from 5 K to 300 K. An a.c. charge current was applied to the injector Pt wire with Keithley 6221 and the voltage across the detector Pt wire was recorded with a lock-in amplifier (NF 5640). The typical a.c. charge current is 100 μA (root mean squared) in amplitude and 3.423 Hz in frequency.

**Acknowledgements**

We thank S. Maekawa, J. Barker, R. Iguchi, T. Niizeki, and D. Hirobe for discussions, and R. Yahiro for experimental help. This work is a part of the research program of ERATO "Spin Quantum Rectification Project" (No. JPMJER1402) from JST, the Grant-in-Aid for Scientific Research on Innovative Area "Nano Spin Conversion Science" (Nos. JP26103005 and JP26103006) and Grant-in-Aid for Research Activity Start-up (Nos. JP18H05841 and JP18H05845) from JSPS KAKENHI, JSPS Core-to-Core program, the International Research Center for New-Concept Spintronics Devices, World Premier International Research Center Initiative (WPI) from MEXT, Japan, Netherlands Organization for Scientific Research (NWO) and NanoLab NL, K.O. acknowledges support from GP-Spin at Tohoku University.

**Author contribution**

K.O. designed the experiment, fabricated the samples, collected all of the data. S.T. formulated the theoretical model. K.O. and S.T. analyzed the data. S.T. and K.O. estimated the parameters. L.J.C. and J.S. carried out the numerical simulation. T.K., B.J.v.W., G.E.W.B., and E.S. developed the explanation of the experimental results. E.S. supervised the project. K.O., S.T., L.J.C., J.S., S.D., T.K., G.E.W.B, B.J.v.W, and E.S. discussed the results and commented on the manuscript.

**Competing Interests**

The authors declare no competing financial interests.

**Correspondence** Correspondence and requests for materials should be addressed to K. O. (email: k.oyanagi@imr.tohoku.ac.jp).

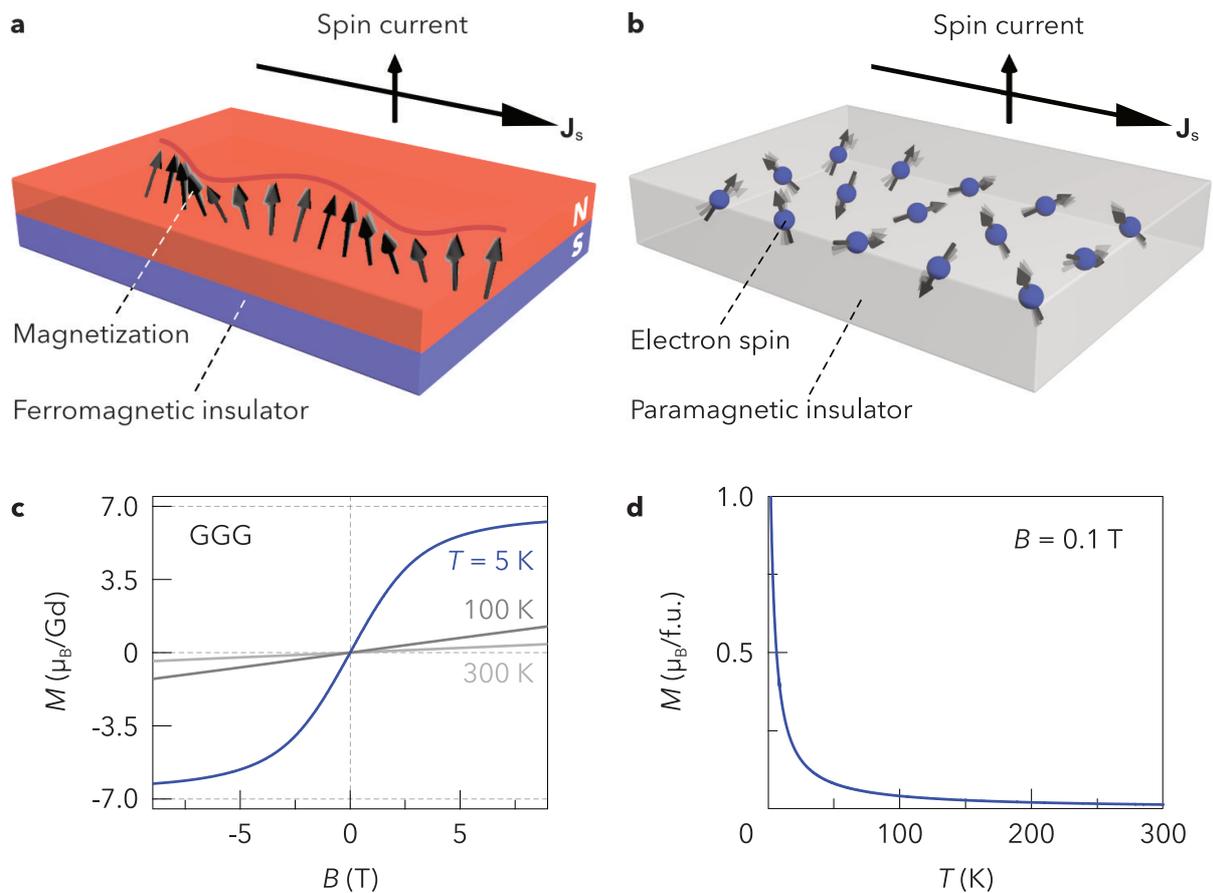

**Figure 1 | Concepts of spin current in a ferromagnetic insulator and a paramagnetic insulator, and paramagnetism of $Gd_3Ga_5O_{12}$. a,** A schematic illustration of a ferromagnet, in which spins are aligned to form long-range order due to strong exchange interaction. **b,** A schematic illustration of a paramagnet, in which directions of localized spins are random due to thermal fluctuations. **c,** Magnetization $M$ as a function of the applied magnetic field $B$ at 5 K, 100 K, and 300 K. The saturation magnetization of GGG is ~ 7 $\mu_B/Gd^{3+}$ at 5 K. **d,** The temperature dependence of the magnetization of GGG at $B = 0.1$ T.

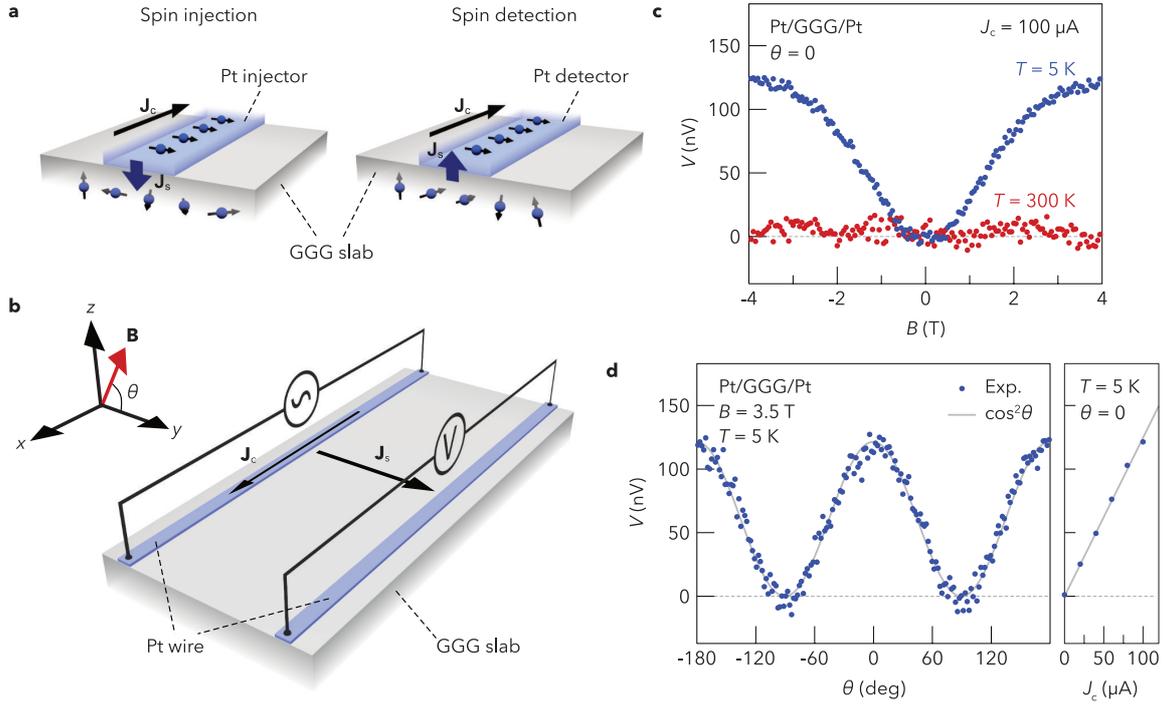

**Figure 2 | Observation of long-range spin transport through a paramagnetic insulator. a,** Schematics of spin injection (left panel) and detection (right panel) at two Pt/GGG contacts. $\mathbf{J}_c$ and $\mathbf{J}_s$ denote the spatial directions of charge and spin currents, respectively. $\mathbf{J}_s$ is injected into GGG by applying $\mathbf{J}_c$ via the SHE in Pt. At the detector, $\mathbf{J}_s$ is driven in the direction normal to the interface and is converted into $\mathbf{J}_c$ via the ISHE in Pt. **b,** A schematic of the experimental set-ups. The nonlocal device consists of two Pt wires patterned on a GGG slab. $\mathbf{B}$ and $\mathbf{E}_{ISHE}$ denote the directions of the applied magnetic field and the electric field induced by the ISHE, respectively. We apply $\mathbf{J}_c$ to the left Pt wire and detect the voltage $L\mathbf{E}_{ISHE}$ between the ends of the right Pt wire with the length $L$. **c,** The $B$ dependence of $V$ at $\theta = 0$ for Pt contacts separated by $d = 0.5$ µm at 300 K (red plots) and 5 K (blue plots). **d,** The $\theta$ and $J_c$ dependence of $V$ for the same device at 5 K. The left panel shows the $\theta$ dependence of $V$ while $B = 3.5$ T was rotated in the $z$-$y$ plane, where the gray line is a $\cos^2\theta$ fit.

The right panel shows the $J_c$ dependence of $V_{max}$, determined by fitting $V_{max}\cos^2\theta$ to the $\theta$ dependence. We subtracted a constant offset voltage $V_{offset}$ from $V$ in **c** and **d** (see Supplementary Information Section B).

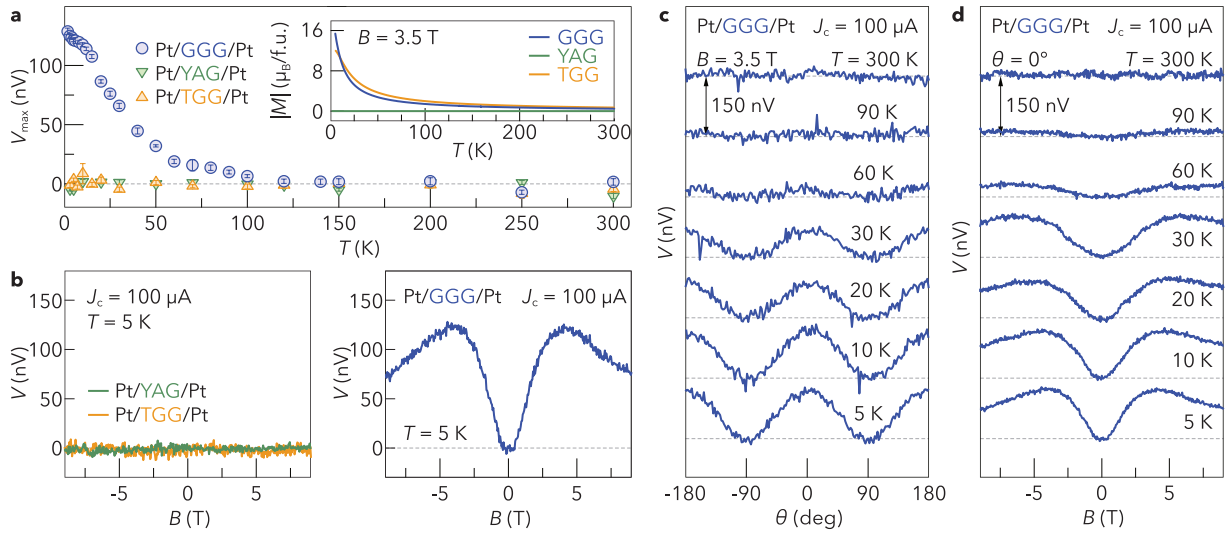

**Figure 3 | Temperature and magnetic field dependence of the nonlocal voltage signal.** All experimental data were obtained by the same device ($d = 0.5$ μm) with a current amplitude of 100 μA. **a,** The temperature ($T$) dependence of the amplitude of the maximum non-local voltage $V_{max}$ for Pt/GGG/Pt, Pt/TGG/Pt, and Pt/YAG/Pt obtained from a sinusoidal fit to the magnetic-field angle $\theta$ dependences of $V$ at $B = 3.5$ T. The error bars represent the 68 % confidence level ($\pm$ s.d.). The inset shows the $T$ dependence of the magnetization $M$ of GGG, TGG, and YAG at $B = 3.5$ T. **b,** Comparison between $V$ for Pt/YAG/Pt, Pt/TGG/Pt, and Pt/GGG/Pt. The left (right) panel shows the $B$ dependences of $V$ for Pt/YAG/Pt and Pt/TGG/Pt (Pt/GGG/Pt) at 5 K. **c,d,** The $\theta$ and $B$ dependence of $V$ for Pt/GGG/Pt at various temperatures. We varied $\theta$ by rotating the field at $B = 3.5$ T in the $z$-$y$ plane. The field was changed from $-9$ T to 9 T at $\theta = 0$ for the $B$ dependence.

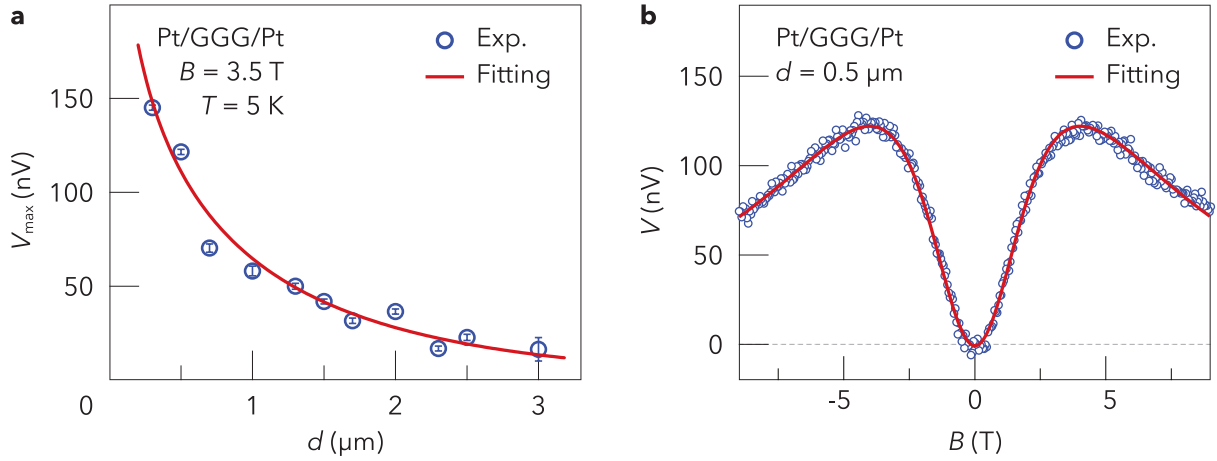

**Figure 4 | Comparison with theory. a,** The experimental results (blue circles) and calculations (solid red line) for $V_{max}$ in Pt/GGG/Pt as a function of the separation $d$ between the Pt contacts and an applied current of 100 µA. We obtain $V_{max}$ by a $\cos^2\theta$ fit to the magnetic-field angle $\theta$ dependence of $V$ at $B = 3.5$ T and $T = 5$ K. By using equation (1), the spin diffusion length $\lambda_{GGG} = 1.82 \pm 0.19$ µm. The error bars represent the 68 % confidence level ($\pm$ s.d.). **b,** The experimental results (blue circles) and calculation (solid red line) of the $B$ dependence of the nonlocal $V$ for Pt/GGG/Pt. We measured the signal in the $d = 0.5$ µm device at 5 K with applying $B$ at $\theta = 0$.